Technical brief

Improvements and simplifications in in-gel fluorescent detection of proteins using ruthenium II tris-(bathophenanthroline disulfonate), the poor man's fluorescent detection method.

Catherine Aude-Garcia, Véronique Collin-Faure, Sylvie Luche, Thierry Rabilloud

CEA-DSV/iRTSV/LBBSI, Biophysique et Biochimie des Systèmes Intégrés, CEA-Grenoble, 17 rue des martyrs, F-38054 GRENOBLE CEDEX 9, France

Université Joseph Fourier, UMR CNRS-CEA-UJF 5092, CEA-Grenoble, 17 rue des martyrs, F-38054 GRENOBLE CEDEX 9

CNRS UMR5092, Biochemistry and Biophysics of Integrated Systems, CEA Grenoble, iRTSV/LBBSI, 17 rue des martyrs, F-38054 GRENOBLE CEDEX 9

Correspondence to

Thierry Rabilloud, iRTSV/LBBSI
CEA-Grenoble, 17 rue des martyrs,
F-38054 GRENOBLE CEDEX 9
Tel (33)-4-38-78-32-12
Fax (33)-4-38-78-44-99
e-mail: Thierry.Rabilloud@ cea.fr

Abstract

Fluorescent detection of proteins is a popular method of detection allying sensitivity, linearity and compatibility with mass spectrometry. Among the numerous methods described in the literature, staining with ruthenium II tris(bathophenanthroline disulfonate) is particularly cost-effective, but slightly cumbersome owing to difficulties in the preparation of the complex and complexity of staining protocols.

We describe here modifications on both aspects that allow to perform a higher contrast staining and offer a more robust method of complex preparation, thereby maximizing the advantages of the method.

Numerous constraints apply to detection methods used in gel-based proteomics. They should be linear over a wide range, homogeneous from protein to protein, and of course sensitive. In addition to these core features, they should also be mass spectrometry-compatible, user-friendly and cost-effective. Among the variety of methods used for protein detection after gel electrophoresis [1], fluorescent methods offer an interesting compromise, especially for detection linearity [2] and for compatibility with mass spectrometry [3]. Among the wide variety of fluorescent detection methods that have been developed to detect proteins after gel electrophoresis, two have emerged as standards in the field, one using epicocconone [4] (marketed under the trade name deep purple) and one using an undisclosed fluorescent ruthenium complex and marketed under the trade name Sypro Ruby [5, 6]. Almost simultaneously with the description of Sypro Ruby staining, another stain using a published ruthenium complex (ruthenium II tris(bathophenanthroline disulfonate) was described [7], with minimal interference with mass spectrometry and much improved cost efficiency compared to commercial formulations. However, the sensitivity of this stain was moderate, and a major improvement was published a few years later [8]. This improvement resulted, however, in a much longer staining period, extending to almost two days after the end of electrophoresis, thereby decreasing the overall productivity of the proteomics setup. This is why intermediate formulations were developed, claiming for equal sensitivity and much improved speed and simplicity of staining [9].

Furthermore, all the protocols using ruthenium II tris(bathophenanthroline disulfonate) are plagued by difficulties in the preparation of the complex. Indeed this preparation involves both complex formation and reduction from ruthenium III to ruthenium II, as most water-soluble ruthenium salts are ruthenium III salts. In the published protocols, this reduction was achieved either with hypophosphoric acid and sodium hydroxide, or by ascorbic acid and sodium hydroxide. As both redox couples are pH-sensitive, the control of the extent of reduction is difficult to manage in the standard proteomics or biochemistry laboratory, so that preparation of the complex occasionally fails. We therefore decided to revisit both aspects of staining with ruthenium II tris(bathophenanthroline disulfonate), namely complex preparation on the one hand, and staining protocol on the other hand, aiming at more simplicity, robustness and performance.

Ammonium formate is a weak but interesting reducing agent [10]. Its solutions are naturally close to neutral pH, thereby needing no pH adjustment, but it is effective only at relatively high temperatures and the overall reducing power is weak. In order to prepare an efficient ruthenium complex, we prepared a solution containing 20 mM potassium pentachloroaquo ruthenate (Alfa Aesar), 60 mM bathophenanthroline disulfonate, disodium salt (Aldrich) and 400 mM ammonium formate (available as a 10 M titrated stock solution from Fluka). Reduction of ruthenium III to ruthenium II and simultaneous complex formation was achieved either by refluxing the solution for three days, or incubating it at 95°C in an oven for three days. In the latter case, because of gas evolution during reduction, it was not possible to tightly close the vessel containing the solution. Thus, a beaker containing water was placed next to the ruthenium solution in the oven set at 95°C, in order to saturate the oven with vapor and limit evaporation in the ruthenium solution. However, some evaporation was frequently encountered, and compensated for at the end of the preparation by water addition to restore the initial volume.

Although much longer to perform, we found this method to be much more reliable than those using ascorbate, hypophosphite or borohydride as reducing agents. In addition, the complex solution was found to be stable for over two years in the refrigerator, with no decrease in staining efficiency. This method using ammonium formate also allowed to replace potassium pentachloroaquo ruthenate by simple ruthenium III chloride at equal molar concentrations, and the resulting, even brighter complex was called alternate complex (see figures 1 and 2).

As to the staining protocol itself, we first compared the three published protocols. The initial protocol [7] (Figure 1A) was clearly less sensitive than the two improved ones [8, 9] (Figure 1B and 1C), but the background was still unsatisfactory with the published protocols, as can be seen in Figure 1. We therefore decided to combine the staining in acidic solution with the destaining step to reduce background. A significant improvement was noted with classical, acetic acid-based solutions (Figure 1 D), but a further improvement in contrast was obtained when the acetic acid-ethanol solutions described in [8] and [9] were replaced by a phosphoric acid-ethanol solution, both for staining and destaining (Figure 1 E), resulting in the simple protocol described here. This protocols describes as follows:

Step 1: after electrophoresis, fix the gel for 1 hour in 1% phosphoric acid (v/v, starting from commercial 85% phosphoric acid) and 30% ethanol

Step 2: stain overnight with 1µM ruthenium complex in 1% phosphoric acid-30% ethanol

Step 3: destain for 4-6 hours in 1% phosphoric acid-30% ethanol. Shorter destaining (1-2 hours) did not afford sufficient background reduction, while longer destaining (e.g. overnight) reduced the sensitivity. The times indicated here are therefore a good compromise

Step 4: rinse in water (one rinse, 5-15 minutes) prior to imaging. This step is just intended to avoid contaminating the UV table with phosphoric acid and ethanol from the liquid film coming along with the gels

Further tests showed that the performances were maintained by using the complex prepared from RuCl3 in place of the complex prepared from K2RuCl5 (Figure 1F), and that the performances were marginally inferior to those obtained with Sypro Ruby (Figure 1G). However, fluorescent staining with such methods was clearly not as sensitive as a silver staining (Figure 1H)

We then evaluated the different staining methods in the context of two-dimensional gels, where the homogeneity of the stain and interference from carrier ampholytes can be evaluated, and the results are described in Figure 2 and in Supplementary Figure 1 with raw images (i.e. without color inversion and gray level conversion) . Here again, it can be seen that the improved protocol (Figure 2 panels A-D) is faster and/or more sensitive (because of better contrast) than those published previously [8, 9] (Figure 2 panels E-I). We also found during this step of the study that 1% phosphoric acid offers the best signal to noise ratio (Figure 2 panels B and D). Higher concentrations of acid increased the background and thus decreased sensitivity (Figure 2 panels A and C). We also tried to reduce further the phosphoric acid concentration below 1%, but the stain became more erratic and less reproducible.
It is also noteworthy that gels stained by the method of Pluder et al. showed a very high background (Figure 2G), in contrast to what is described in the original publication [9]. However, it shall be mentioned that the original publication uses a laser excitation at 532 nm, i.e. quite far from the excitation maxima of the ruthenium complex (280 and 470 nm), whereas excitation was performed at 312 nm in the present study.

Finally, we evaluated the new staining method in a real proteomic-type experiment based on two-dimensional gels, and the results are shown on Figure 3 and in Supplementary Figure *2* with raw images (i.e. without color inversion and gray level conversion). The detection protocol as a whole, i.e. staining and imaging, was found to be very reproducible, due to its low number of steps and to flexibility in each step duration. As a result, the median rsd (relative standard deviation) in a set of four parallel gels was found to be 15-16%, to be compared to the 20-25% previously described for several detection methods including colloidal Coomassie Blue [13]. We could stain up to four gels per dish, provided that 250ml of solution was present per gel (160x200x1.5 mm) at each step.

Compared to the early protocols [7, 8], the new staining protocol uses far fewer steps, thereby minimizing the risk of external contamination of the protein spots, e.g. by keratins. Avoidance of acetic acid also decreases the risk of artefactual acetylation.

In conclusion, the modified protocol described in this technical brief offers an improved detection method using the ruthenium bathophenanthroline complex, being faster and simpler, and offering better contrast than previously published protocols using the same fluorescent agent. Furthermore, an improved preparation of the fluorescent complex that can be made in all biochemical laboratories without any special glassware is also described. This protocol offers the benefits of the metal-based fluorescent complex detection, i.e. sensitivity, robustness, resistance to fading, at a minimal cost.

References

[1] Miller, I., Crawford, J., Gianazza, E., Protein stains for proteornic applications: Which, when, why? *Proteomics* 2006, *6*, 5385-5408.
[2] Patton, W. F., A thousand points of light: The application of fluorescence detection technologies to two-dimensional gel electrophoresis and proteomics. *Electrophoresis* 2000, *21*, 1123-1144.
[3] Chevalier, F., Centeno, D., Rofidal, V., Tauzin, M*., et al.*, Different impact of staining procedures using visible stains and fluorescent dyes for large-scale investigation of proteomes by MALDI-TOF mass spectrometry. *Journal of Proteome Research* 2006, *5*, 512-520.
[4] Mackintosh, J. A., Choi, H. Y., Bae, S. H., Veal, D. A*., et al.*, A fluorescent natural product for ultra sensitive detection of proteins in one-dimensional and two-dimensional gel electrophoresis. *Proteomics* 2003, *3*, 2273-2288.
[5] Berggren, K., Chernokalskaya, E., Steinberg, T. H., Kemper, C*., et al.*, Background-free, high sensitivity staining of proteins in one- and two-dimensional sodium dodecyl sulfate-polyacrylamide gels using a luminescent ruthenium complex. *Electrophoresis* 2000, *21*, 2509-2521.
[6] Steinberg, T. H., Chernokalskaya, E., Berggren, K., Lopez, M. F*., et al.*, Ultrasensitive fluorescence protein detection in isoelectric focusing gels using a ruthenium metal chelate stain. *Electrophoresis* 2000, *21*, 486-496.
[7] Rabilloud, T., Strub, J. M., Luche, S., van Dorsselaer, A., Lunardi, J., Comparison between Sypro Ruby and ruthenium II tris (bathophenanthroline disulfonate) as fluorescent stains for protein detection in gels. *Proteomics* 2001, *1*, 699-704.
[8] Lamanda, A., Zahn, A., Roder, D., Langen, H., Improved Ruthenium II tris (bathophenantroline disulfonate) staining and destaining protocol for a better signal-to-background ratio and improved baseline resolution. *Proteomics* 2004, *4*, 599-608.
[9] Pluder, F., Beck-Sickinger, A. G., One-step procedure for staining of proteins with ruthenium II tris (bathophenanthroline disulfonate). *Analytical Biochemistry* 2007, *361*, 299-301.
[10] Andrade, C. K. Z., Silva, W. A., One-step reduction of chalcones to saturated alcohols by ammonium formate/palladium on carbon: A versatile method. *Letters in Organic Chemistry* 2006, *3*, 39-41.
[11] Chevallet, M., Luche, S., Rabilloud, T., Silver staining of proteins in polyacrylamide gels. *Nat Protoc* 2006, *1*, 1852-1858.
[12] Tastet, C., Lescuyer, P., Diemer, H., Luche, S*., et al.*, A versatile electrophoresis system for the analysis of high- and low-molecular-weight proteins. *Electrophoresis* 2003, *24*, 1787-1794.
[13] Luche, S., Lelong, C., Diemer, H., Van Dorsselaer, A., Rabilloud, T., Ultrafast coelectrophoretic fluorescent staining of proteins with carbocyanines. *Proteomics* 2007, *7*, 3234-3244.

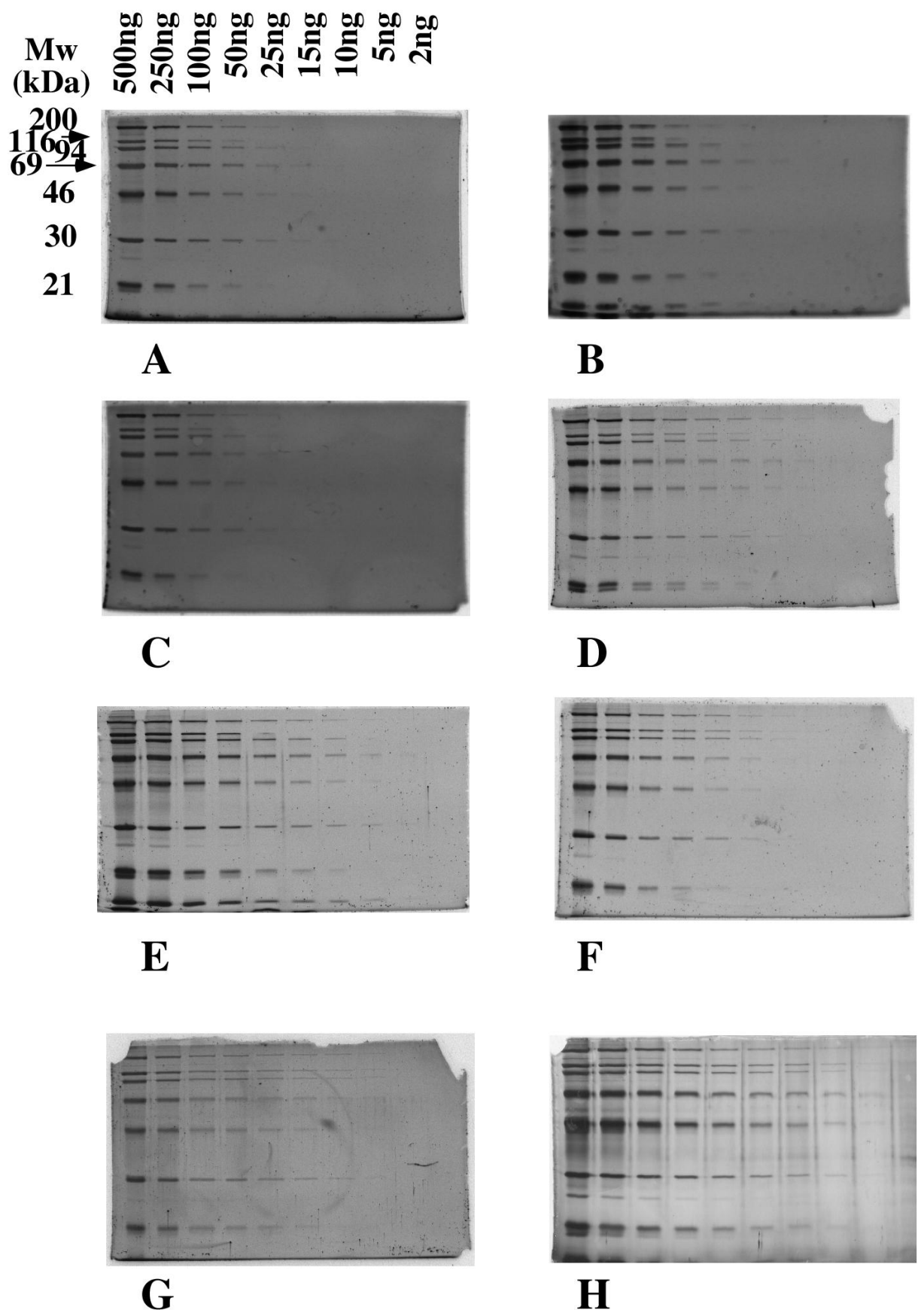

Figure 1: sensitivity evaluation with 1D SDS PAGE

Molecular weight marker proteins (from Bio-Rad, broad range) were diluted to obtain the following loads per protein and per lane lane 1: 500 ng; lane 2: 250 ng; lane 3: 100 ng; lane 4: 50 ng; lane 5: 25 ng ; lane 6: 15 ng; lane 7: 10 ng; lane 8: 5 ng; lane 9: 2 ng.

The proteins were loaded on top of 12% acrylamide gels ( 80 x 50 x 1mm) run in Tris-glycine buffer.
After electrophoresis, the proteins were detected by fluorescence, and imaged on a UV table (312nm emission) with a digital camera (canon A95) equipped with a UV filter and a yellow filter to cut the light emission of the table. Photography for 4 seconds at f-stop 8.

A: staining by 200 nM complex in water, no destaining [7].
B: staining by 1 µM complex in water, destaining in 10% acetic acid+40% ethanol[8].
C: staining by 1 µM complex in 10% acetic acid+30% ethanol, no destaining [9].
D: staining by 1 µM complex in 10% acetic acid + 30 % ethanol, destaining in the same solution.
E: staining by 1 µM complex in 2% phosphoric acid + 30 % ethanol, destaining in the same solution
F: staining by 1µM alternate complex (from RuCl3) in 2% phosphoric acid + 30 % ethanol, destaining in the same solution
G: staining with Sypro Ruby [5]
H: silver staining (fast silver nitrate staining) [11]

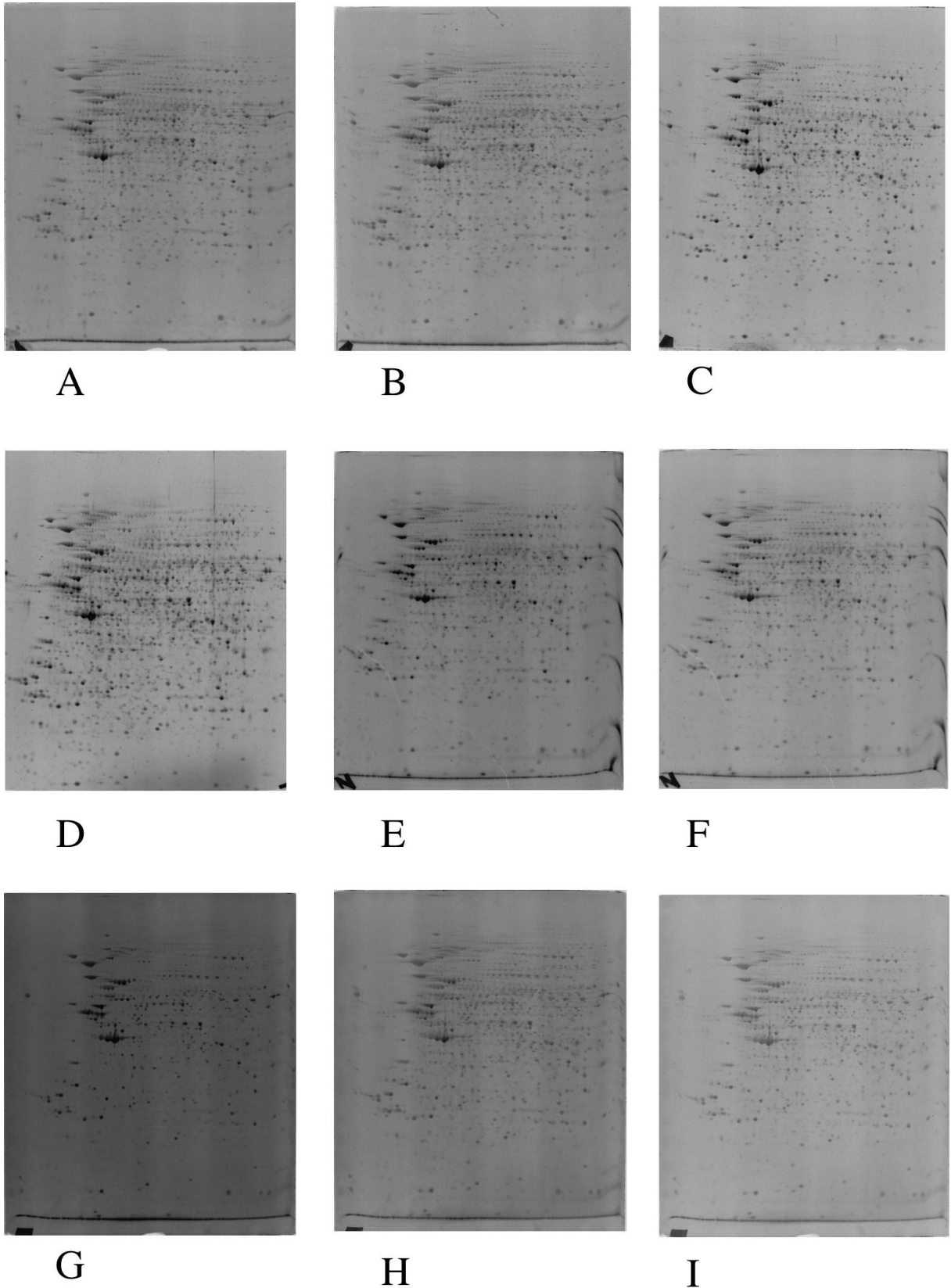

Figure 2: sensitivity evaluation with 2D PAGE

Total protein extracts from the LADMAC cell line (ATCC CRL 2420) were analyzed by 2D electrophoresis. 250 µg were loaded on the first dimension gels (immobilized pH 4-8 gradients)

The strips were transferred on top of 10% acrylamide gels ( 80 x 50 x 1mm) run in Tris-taurine buffer.[12]
After electrophoresis, the proteins were detected by fluorescence, and imaged on a UV table (312nm emission) with a digital camera (canon A95) equipped with a UV filter and a yellow filter to cut the light emission of the table. Photography for 4 seconds at f-stop 8, unless otherwise indicated.

A: staining by 1 µM complex in 2% phosphoric acid + 30 % ethanol, destaining in the same solution
B: staining by 1 µM complex in 1% phosphoric acid + 30 % ethanol, destaining in the same solution
C: staining by 1 µM alternate complex in 2% phosphoric acid + 30 % ethanol, destaining in the same solution
D: staining by 1 µM alternate complex in 1% phosphoric acid + 30 % ethanol, destaining in the same solution
E: staining by the method of Lamanda et al [8]
F: same as E, but photography for 2 seconds
G: staining by the method of Pluder et al.[9]
H: same as G, but photography for 2 seconds
I: same as G, but photography for 1 second

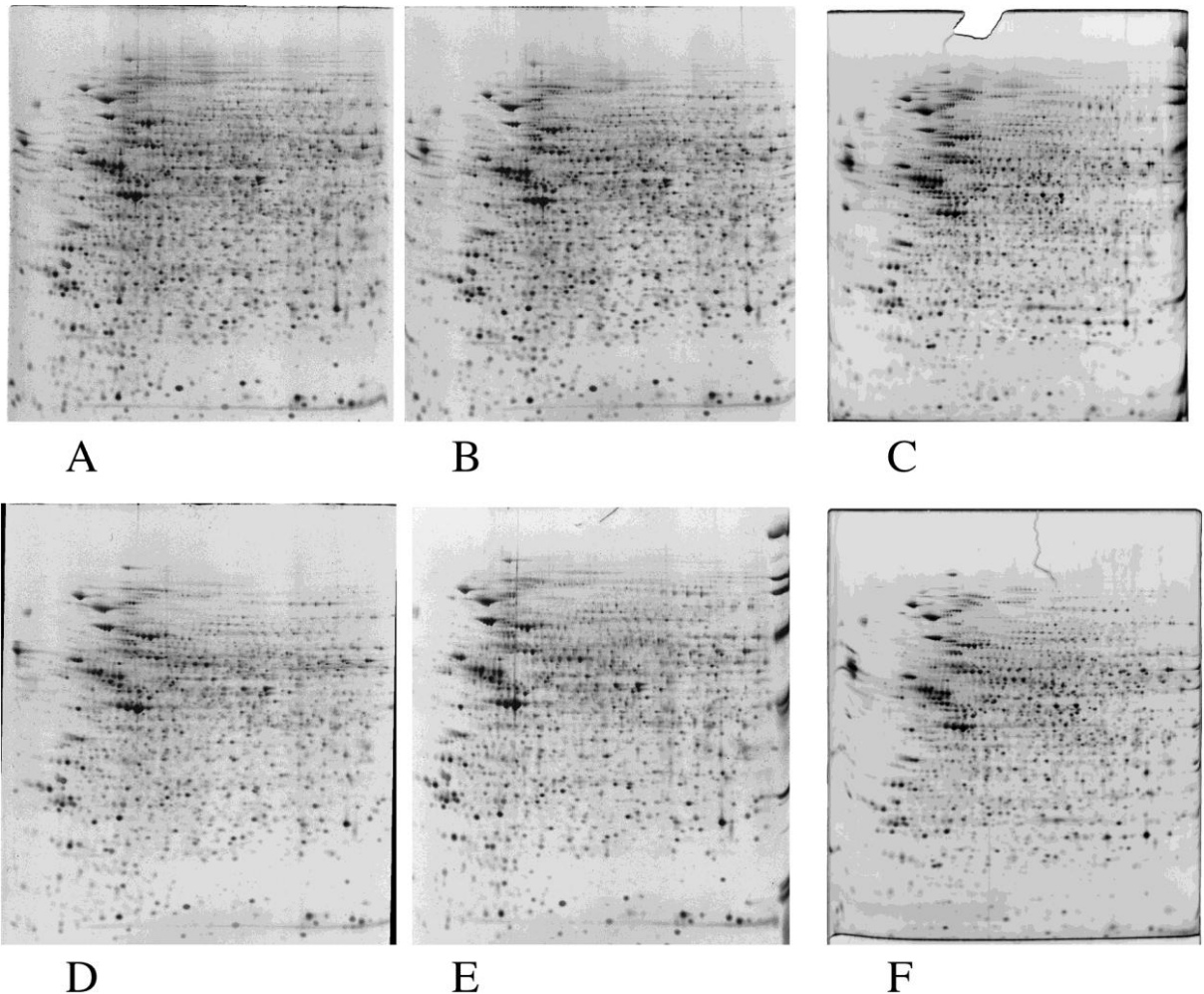

Figure 3: reproducibility analysis in a proteomics context
Total protein extracts from the M1 myeloblastoid cell line (ATCC TIB 192), induced or not with dexamethasone for 3 days, were analyzed by 2D electrophoresis. 300 µg were loaded on the first dimension gels (immobilized pH 4-8 gradients). The second dimension was 10% acrylamide gels, operating in the Tris Taurine system [12]. Staining by 1µM complex in 1% phosphoric acid + 30 % ethanol, destaining in the same solution, as described in the main text.
Imaging on a UV table (312nm emission) with a digital camera (canon A95) equipped with a UV filter and a yellow filter to cut the light emission of the table. Photography for 4 seconds at f-stop 8.
The images were then analyzed by the Delta2D software (v 3.6, Decodon, Germany). For each group (control and treated cells) ca. 1600 spots were detected, and the median rsd (relative standard deviation) was calculated. The control group had a median rsd of 15.1 % and the treated group had a median rsd of 15.9 %. Two gels are shown for each group, while four were made and included per group for the computerized analysis.

A, B: control cells, fluorescence detection 300 µg/gel. C: control cells, 120 µg loaded, silver staining
D, E: dexamethasone-treated cells, fluorescence detection 300 µg/gel. F: dexamethasone-treated cells, 120 µg loaded, silver staining